\documentclass[epsfig]{mn2e}
\usepackage{epsf}
\usepackage{epsfig}
\usepackage{graphicx}
\usepackage{xspace}
\usepackage{subfigure}

\newcommand{\eg}{{\sl e.g.~}}
\newcommand{\ie}{{\sl i.e.~}}
\newcommand{\etal}{{et al.~}}
\def\simlt{\lower.5ex\hbox{$\; \buildrel < \over \sim \;$}}
\def\simgt{\lower.5ex\hbox{$\; \buildrel > \over \sim \;$}}
\def\simpropto{\lower.2ex\hbox{$\; \buildrel \propto \over \sim \;$}}

\newcommand{\degs}{$^{\circ}$}
\newcommand{\pms}{$\pm$}

\newcommand{\ergs}{$\,$erg$\,$s$^{-1}$}

\newcommand{\apj}{ApJ}
\newcommand{\apjs}{ApJS}
\newcommand{\apjl}{ApJ}
\newcommand{\mnras}{MNRAS}

\newcommand{\nat}{Nature}
\newcommand{\aap}{A\&Ap}
\newcommand{\prd}{Phys.~Rev.~D}

\def\simlt{\lower.5ex\hbox{$\; \buildrel < \over \sim \;$}}
\def\simgt{\lower.5ex\hbox{$\; \buildrel > \over \sim \;$}}

\begin{document}

\title[511 keV Emission in the Galactic Centre]
     {On the Origin of the 511 keV Emission in the Galactic Centre}
\author[Bandyopadhyay, Silk, Taylor \& Maccarone]
{Reba M. Bandyopadhyay$^{1}$\thanks{email: {\tt reba@astro.ufl.edu}},
  Joseph Silk$^{2}$, James E. Taylor$^{3}$ and Thomas J. Maccarone$^ 
{4}$\\
$^{1}$Department of Astronomy, 211 Bryant Space Science Centre,  
Gainesville, FL 32611-2055 USA\\
$^{2}$Department of Astrophysics, Denys Wilkinson Building, 1 Keble Road, Oxford OX1 3RH, UK\\
$^{3}$Department of Physics \& Astronomy, 200 University Avenue West,  
Waterloo, Ontario, N2L 3G1 Canada\\
$^{4}$School of Physics \& Astronomy, University of Southampton,  
Highfield, Southampton SO17 1BJ, UK}
\date{\today}
\pubyear{2007}


\maketitle

\begin{abstract}
Diffuse 511 keV line emission, from the annihilation of cold
positrons, has been observed in the direction of the Galactic Centre
for more than 30 years. The latest high-resolution maps of this
emission produced by the SPI instrument on INTEGRAL suggest at least
one component of the emission is spatially coincident with the
distribution of $\sim$70 luminous, low-mass X-ray binaries detected in
the soft gamma-ray band. The X-ray band, however, is generally a more
sensitive probe of X-ray binary populations. Recent X-ray surveys of
the Galactic Centre have discovered a much larger population ($>$4000)
of faint, hard X-ray point sources. We investigate the possibility
that the positrons observed in the direction of the Galactic Centre
originate in pair-dominated jets generated by this population of {\it
fainter} accretion-powered X-ray binaries.  We also consider briefly
whether such sources could account for unexplained diffuse emission
associated with the Galactic Centre in the microwave (the WMAP `haze')
and at other wavelengths. Finally, we point out several unresolved
problems in associating Galactic Centre 511 keV emission with the
brightest X-ray binaries.
\end{abstract}



\section{Introduction}\label{sec:1}

Emission at 511 keV, the characteristic signature of positron
annihilation, has been observed in the direction of the Galactic
Centre (GC), since the 1970s.  Apparently diffuse gamma-ray emission
at approximately this energy was first detected in 1970 by the
balloon-borne experiments of the Rice group (Johnson, Harnden, \&
Haymes 1972; Johnson \& Haymes 1973; Haymes \etal 1975; preliminary
indications were also reported in Haymes \etal 1969) and was confirmed
as positron annihilation emission in 1978 by the balloon-borne
experiments of the Bell-Sandia group (Leventhal, MacCallum, \& Stang
1978; Leventhal \etal 1980).  High-energy balloon experiments and
space observatories through the early 1990s continued to detect the
511 keV emission; however the relatively low spatial resolution of
these detectors prevented determination of the true location and
distribution of the emission (see Purcell \etal 1997, Jean \etal 2003,
or Teegarden \etal 2005 for a summary of these early observations).
Specifically, from these data it is unclear whether the emission is
truly diffuse, or if it originates either from a single discrete
source (\eg Sgr A* or 1E1740.7-2942,``The Great Annihilator'') or from
a small number of discrete but unresolved sources. Some early
detections suggested time variability in the signal, indicating a
small number of discrete sources, but these variations were not
confirmed in subsequent observations (\eg Purcell \etal 1997;
Teegarden \etal 2005).

With the advent of space observatories of increasing sensitivity,
spectral coverage, and spatial resolution, the resultant improvement
in data quality now provides much stronger constraints on the origin
of the emission.  In particular, observations of the GC by the SPI
spectrometre on the satellite INTEGRAL (the INTErnational Gamma-Ray
Astrophysicsl Laboratory) have recently produced the most detailed map
of the anomalous 511 keV emission to date (Kn{\"o}dlseder \etal 2005;
Weidenspointner \etal 2006, 2007).  The SPI/INTEGRAL map clearly shows
this emission arising from the central $\sim$1.5 kpc ($l < 10$\degs)
of the Galaxy, with a fainter component of 511 keV flux detected from
the remainder of the Galactic disk. The smoothness of the emission
constrains the number of discrete sources responsible to be in excess
of $\sim$ 8 (Kn{\"o}dlseder \etal 2005), as do point source limits of
$1.6 \times 10^{-4}$ ph cm$^2$ s$^{-1}$ from searches with the IBIS
imager on INTEGRAL (de Cesare \etal 2006). Most recently,
Weidenspointner \etal (2008) have detected a longitudinal asymmetry in
the disk component of the emission, which matches the asymmetry in the
distribution of the brightest low-mass X-ray binaries (LMXBs) detected
by the IBIS instrument on INTEGRAL, suggesting that these $\sim70$
objects may account for much of the disk component of the 511 keV
emission.

Diffuse emission from the Galactic Centre, unaccounted for by known
sources, has also been observed at several other wavelengths, notably
high-frequency radio (21--63 GHz) with WMAP (Finkbeiner 2004a) and
soft X-rays (1--10 keV) with \eg HEAO and Chandra (Worrall \etal 1982,
Muno \etal 2004).  The origin and nature of the diffuse emission is
puzzling in each case.  In the case of the diffuse X-ray component, a
population of discrete but unresolved point sources was originally
postulated as a possible source of the emission (see \eg Skibo \etal
1997, Valinia \& Marshall 1998).

With the advent of high-resolution X-ray imaging using Chandra, a
large new population of low X-ray luminosity sources has indeed been
identified in the central 2\degs$\times$0.8\degs of the GC
(300$\times$120 pc at a GC distance of 8.5 kpc; Wang \etal 2002, Muno
\etal 2003, Muno \etal 2006).  Many of these $\sim$4200 discrete X-ray
sources are likely to be accreting binaries, including high and low
mass X-ray binaries and cataclysmic variables.  However, it is worth
noting that these detected sources only account for 10\% of the
previously observed diffuse 2-10 keV emission (Ebisawa \etal 2001,
Muno \etal 2003), so an even larger population of still fainter
sources (with $L_{X} < 10^{31}$\ergs) could exist at the GC (but see
Muno \etal 2004 who argue that the spectral characteristics of the
remaining diffuse X-ray emission are inconsistent with a stellar X-ray
source origin).  In any case, it is not clear how much these new
populations might contribute to the diffuse flux at lower (\eg WMAP)
or higher (\eg INTEGRAL) energies.

In this paper, we consider the hypothesis that jet outflows from X-ray
binary systems (XRB), which are accretion-powered mass-transferring
binaries containing a black hole (BH) or neutron star (NS), are the
main source of the 511 keV emission in the GC. This idea has been
explored previously by various authors, notably Ramaty \& Lingenfelter
(1979), Prantzos (2004), Kn{\"o}dlseder \etal (2005), and Guessoum
\etal (2006).  These previous studies generally focused on the
large-scale discrete jet ejections produced by the class of luminous
black hole XRBs known as ``microquasars''.  However, it is now thought
that luminous black hole XRBs much more commonly produce
lower-luminosity ``steady'' jets, and that these outflows are ``on''
for a substantially greater fraction of the XRB duty cycle than the
large-scale ejection events (Gallo \etal 2006).  We note that neutron
star X-ray binaries are, in principle, also possible sources of jet
positrons.  The fraction of low magnetic field NS XRBs with outflows
could be as high as $\sim$ 100\% (Fender 2006; Migliari \& Fender
2006).  However, they show a different scaling than black hole XRBs
between X-ray luminosity and radio luminosity (i.e. $L_R \propto
L_X^{0.7}$ in BH systems and $L_R \propto L_X^{1.4}$ in NS systems;
Miglari \& Fender 2006).  As a result of this scaling, the jets from
quiescent neutron star X-ray binaries are likely to be several orders
of magnitude weaker than those from quiescent BH X-ray binaries, if
one assumes the same scaling relations continue into quiescence.  More
likely, though, is that the scaling relation becomes even steeper for
neutron stars as they fade more deeply into quiescence than the
current radio flux limits allow us to probe; the emission from the
faintest neutron star XRBs is generally dominated by thermal crustal
emission from cooling neutron stars, rather than by accretion power
(\eg Rutledge \etal 2001).

Also, while some luminous high-mass X-ray binaries (HMXBs) have been
observed to emit jets (e.g. Cyg X-1, Cyg X-3), these systems are
predominantly located in the Galactic disk (Grimm \etal 2002).  Thus
these canonical BH HMXBs -- of which there are only a few known in the
entire Milky Way -- are not a class of objects which can cause an
excess of high-energy flux in the Bulge relative to the Galactic disk.
Recent INTEGRAL observations have detected a population of
lower-luminosity, X-ray hard, high-mass XRBs including highly-obscured
HMXBs and ``supergiant fast X-ray transients'', a number of which are
located in the Bulge (Chaty \etal 2008).  However, there is as yet no
evidence for jet outflows in these HMXBs.  Furthermore, the nature of
the compact objects in these systems is currently not known; but we
note that of those canonical HMXBs for which the nature of the compact
object has been identified, the majority contain neutron stars rather
than black holes.  Finally, in a detailed study of the spatial
distribution of INTEGRAL-detected XRBs (including the new
highly-obscured systems), Lutovinov \etal (2005) find that the angular
distribution of HMXBs in the inner Galaxy is significantly different
from that of LMXBs -- specifically, LMXBs are clearly the dominant
population within the Bulge, significantly overabundant as compared
with the HMXBs in this region.  Thus while we cannot rule out a
high-energy flux contribution in the Bulge from these newly discovered
low-luminosity HMXBs, we will not consider them further here.

Therefore in this paper we primarily consider jets from low-mass black
hole binaries.  It seems plausible that steady, low-luminosity
outflows from Galactic BH XRBs contribute substantially to the
annihilation line emission, and possible also to the diffuse emission
observed at other wavelengths.

The outline of the paper is as follows. In section 2, we first review the 
511 keV observations of the GC and discuss possible sources for the 
positrons. In section 3, we outline a simple model for positron production in
(quiescent) LMXB jets.  In section 4, we briefly examine the unexplained
diffuse GC emission observed at microwave and X-ray wavelengths, 
and consider whether this emission could also originate in outflows from
low-luminosity LMXBs. In section 5, we discuss the recent association of 
the disk component of the 511 keV emission with bright LMXBs detected 
by IBIS, pointing out some unresolved problems with this association. We 
conclude by proposing several observational tests that may help elucidate 
the nature of the population responsible for the 511 keV emission.

\section{Unexplained Emission from the Galactic Centre}\label{sec:2}

\subsection{Observations}

Since its initial discovery almost 40 years ago, our picture of the
annihilation emission from the GC has gradually become clearer. The
emission amounts to $\sim10^{-3}$ photons cm$^{-2}$ s$^{-1}$, coming
from a region roughly $10^{\circ}$ in radius around the GC. Assuming
the mean distance to the positron sources is the distance to the GC,
8.5 kpc, this flux corresponds to an integrated luminosity of
$10^{43}$ photons/s emitted within a region 1.5 kpc in radius. After
initial suggestions of temporal variability in the flux, extensive
observations over the 1990s have ruled this out at any substantial
level (see Purcell \etal 1997 for a summary of this evidence).

The detailed spatial distribution of the emission has become much
clearer since observations first by OSSE on CGRO, and then more
recently by SPI on INTEGRAL. The most recent analyses of 4 years of
SPI/INTEGRAL data (Weidenspointner \etal 2008) suggest two main
components: a central bulge and an asymmetric disk. The bulge
component is reasonably well described by a single Gaussian with a
FWHM of 6$^\circ$, but even better described by a superposition of two
Gaussians of FWHM of 3$^\circ$ and 12$^\circ$, or alternately a
compact, symmetric bulge component from $R = 0$--0.5 kpc, and an
extended shell of emission from $R = 0.5$--1.5 kpc. The disk component
is now detected at $\sim14\sigma$ in 4 years of SPI, and appears to be
be asymmetric at 3.8$\sigma$ significance (Weidenspointner \etal
2008), with 1.8 times more emission at negative longitudes than at
positive ones. The total flux from the disk is $\sim7\times 10^{-4}$
photons cm$^{-2}$ s$^{-1}$, \ie comparable to the bulge flux.  There
is also marginal evidence for emission more than 1.5 kpc from GC
(Kn{\"o}dlseder \etal 2005), possibly extending out as much as
40$^\circ$/5 kpc (Bouchet \etal 2008), though detection here is
hampered by the very uneven exposure maps of INTEGRAL away from the
GC. On the other hand there has been no significant detection of the
emission more than 40--50 degrees away from the GC (Teegarden \etal
2005; Weidenspointner \etal 2007).

Detailed analysis of the annihilation line profile, the positronium
continuum below 511 keV, and the spectrum of the diffuse gamma-ray
background at higher energies place further limits on the origin of
the line. INTEGRAL and COMPTEL measurements of the diffuse background
at 1--10 MeV indicates that the positrons must be produced at
relatively low energy, since otherwise inflight annihilation of
relativistic positrons off background electrons would produce a
visible bump in the spectrum.  This constraint on the injection
energy, first identified by Agaronyan \& Atoyan (1981), places an
upper limit of $\sim$3 MeV on the mass of a light dark matter
candidate producing positrons through annihilation in a neutral medium
(Beacom \& Y\"uksel 2006; note that in this case internal
bremsstrahlung in the annihilation process also contributes to the
high-energy emission). Allowing for a partially ionized medium (and
for production mechanisms without associated internal bremsstrahlung
emission), the more general limit on the injection energy is
$\sim$4--8.5 MeV (Sizun, Cass{\'e}, \& Schanne 2006, 2007).
Furthermore, low-energy positrons rarely annihilate directly; instead
the majority form positronium, a short-lived bound system composed of
an electron-positron pair. Positronium then decays either from the
singlet (``para-positronium'', spins anti-parallel) state, via the
emission of two photons at 511 keV, or from the triplet
(``ortho-positronium'', spins parallel) state by the emission of three
photons, leading to a continuum below 511 keV.  Detailed analysis of
the line profile and line-to-continuum ratio can constrain the
properties of the medium in which the positrons propagate and
annihilate. The recent analyses of Churazov \etal (2005) and Jean
\etal (2006) indicate that the dominant emission source cannot be
located in the very hot or very cold components of the ISM, but that
annihilation probably occurs in the warm neutral and ionized medium.

Given these constraints and the expected lifetimes for positrons in
different components of the ISM, the implication is that positrons are
either produced in, and annihilate in, the warm medium, or are
produced in the hot medium but propagate further and annihilate in the
warm medium at the edge of hot bubbles. In the former case they would
travel only 50--100 pc ($\sim0.5^\circ$ at GC) from their source
before annihilating, and in even in the latter they would
annihilate within $\sim250$ pc of their source (Jean \etal
2006). Thus, in the absence of large-scale magnetic fields (which
could cause the positrons to propagate even further from their
sources; Prantzos 2006) the spatial extent of the observed emission
implies sources distributed over a similar area on the sky.

\subsection{Possible Positron Sources}

Since its discovery, many possible sources have been invoked to
explain the 511 keV emission from the Galactic Centre. A number that
were initially consistent with early detections are now disfavoured,
given more accurate maps and flux measurements. Some sources, \eg
cosmic rays interacting with the intergalactic medium (Ramaty \etal
1970; Lingenfelter \& Ramaty 1982), or populations of young stellar
objects such as pulsars (Sturrock 1971), or supernovae or Wolf-Rayet
stars (via the radioactive nuclei they produce -- Clayton 1973; Ramaty
\& Lingenfelter 1979; Signore \& Vedrenne 1988; Woosley \& Pinto 1988;
Lingenfelter \& Ramaty 1989; Milne \etal 2002), would produce a more
flattened distribution on the sky, inconsistent with the bulge
component. Some other possible candidates, \eg Sgr A* (Lingenfelter \&
Ramaty 1982; Rees 1982; Ozernoy 1989; Ramaty \etal 1992), a single GRB
or hypernova at Galactic Centre (Lingenfelter \& Hueter 1984), or a
population of classical novae (Purcell \etal 1997), should appear as a
single point source and/or show temporal variability (although models
where stronger emission of positrons from Sgr A* occured at some time
in the past may be possible -- see Cheng, Chernyshov \& Dogiel 2006
and 2007; and Totani 2006).  While some of these sources may account
for part of the annihilation flux (\eg the radioactive decay of
$^{26}$Al from young massive stars should account for 20-30\% of the
511 keV emission in the central part of the Galactic disk; Bouchet
\etal 2008), the majority of the 511 keV flux now seems most likely to
come from a distributed population of faint sources that traces the
old stellar bulge. We refer the reader to \eg Kn{\"o}dlseder \etal
(2005) for further discussion of the alternatives.

Of sources associated with old stellar populations, type Ia SNe are
too rare to account for much of the flux. Hypernovae/GRB progenitors
(Cass{\'e} \etal 2004) or GRBs (Parizot \etal 2005; Bertone \etal
2006) might produce enough positrons, but it is not clear they could
necessarily reproduce the smoothly distributed emission seen by
SPI. Another possibility worth mentioning is pulsars. Rotationally
powered pulsars are well known to produce high energy particles. In
recent years, it has become clear through the observation of a
synchrotron nebula around the pulsar 1957+20 (Stappers \etal 2003)
that even millisecond pulsars (MSPs) can have significant
pair-dominated winds.  Given their considerably larger numbers
compared with young pulsars, especially in regions of old stellar
populations like the Galactic Bulge, the pair flux from MSPs should
dominate the total pair flux from pulsars, and it has been shown that
reasonable parameter values for pulsar numbers and pair flux rates can
reproduce the observed annihilation line rate (Wang, Pun \& Cheng
2006).  Thus MSPs are a plausible source of the 511 keV line emission.
However, pulsars are expected to produce very high energy ($>$10 MeV)
positrons; but as noted earlier, the INTEGRAL and COMPTEL measurements
of the diffuse 1-10 MeV background indicate that the positrons
responsible for the observed annihilation line must be produced at
relatively low energy, as otherwise a visible bump in the spectrum
from inflight annihilation of relativistic positrons would be observed
(Agaronyan \& Atoyan 1981).  Thus we will not consider MSPs further in
this paper.

Finally, a large number of more exotic possibilities have been
suggested to explain the 511 keV emission. They include annihilation
of an MeV-scale particle, such as ``light'' dark matter (Boehm \etal
2004); sterile neutrinos or pseudo-scalar relics with masses $< 100$
MeV (Picciotto \& Pospelov 2005); decays of axions (Hooper \& Wang
2004), sub-GeV scale neutralinos (Gunion \etal 2006; Bird \etal 2006),
or other supersymmetric particles (Takahashi \& Yanagida 2006);
Q-balls (Kasuya and Takahashi 2005), mirror matter (Foot \& Silagadze
2005), moduli (Kawasaki \& Yanagida 2005; Kasuya and Kawsaki 2006),
superconducting cosmic strings (Ferrer \& Vachaspati), droplets of
superconducting quark matter (Oaknin \& Zhitnitsky 2004; McNeil, Forbes,
\& Zhitnitsky 2008), TeV-scale particles with an excitation at the MeV
scale (Finkbeiner \& Weiner 2007; Pospelov \& Ritz 2007) or small,
accreting black holes (Titarchuk \& Chardonnet 2006).

\section{A Model for Positron Production in LMXB Jets}

\subsection {Emission from Accreting Binaries}\label{sec:3}

Over the past decade, multiwavelength observations of X-ray binaries
have clearly shown that these systems produce relativistic jet
outflows, analogous to those observed in AGN and quasars (see Fender
2006 for a review).  There are known accreting binary populations in
the GC, and there is increasing evidence that all accreting sources
containing black holes and neutron stars have a {\it compact, steady}
jet outflow during their ``hard'' states, when their X-ray luminosity
is relatively low (indeed, there is evidence that these compact jets
are also present during the ``quiescent'' state when the source X-ray
luminosity is extremely low; Gallo \etal 2006, 2007).  Several
(perhaps all) BH XRBs also have {\it transient} powerful, large-scale,
extended jets, which are associated with the transition between the
``hard'' or ``quiescent'' states and the X-ray luminous ``high/soft''
state.  These large-scale ejection events are episodic but their duty
cycle is poorly known; however, it is clear that most BH XRBs spend
less time in the transitional regime where the large-scale jets are
emitted than in the lower X-ray luminosity hard (or quiescent) states.
We therefore focus here on the low-luminosity steady jets.

Extending the results of deep X-ray surveys of the central Galactic
Bulge (Wang \etal 2002, Muno \etal 2003, Muno \etal 2006) to the
entire Bulge, we would expect roughly $10^5$ X-ray point sources
similar to those detected by the Chandra surveys.  While a large
fraction of these are likely to be cataclysmic variables (see \eg Muno
\etal 2003), a few percent of the sources are likely to be quiescent
low-mass BH X-ray binaries, sources which trace the older stellar
population of the Galaxy and thus are observed to be concentrated in
the inner Galaxy.  Such quiescent systems might be good candidates for
the jet positron source, consistent with known point sources down to
$10^{31}$ergs/s.  These jets could be sources of high-energy electrons
and protons, and also, as they interact with the ISM, of positrons.

Alternatively, this could be done by low energy cosmic ray injection
in jets with associated acceleration to high energies in the jet-ISM
interaction sites (\eg Heinz \& Sunyaev 2003; Fender, Maccarone \& van
Kesteren 2005; Heinz \& Grimm 2005).  These two distinct components
would in turn produce diffuse emission across the electromagnetic
spectrum via positron annihilation, synchrotron emission, and perhaps
heating of the ISM.  This solution is preferable to many of the more
exotic scenarios, which require fine tuning to avoid constraints on
positron production at energies above $\sim$10 MeV, and cannot
simultaneously produce both low-energy positrons and high-energy
electrons and protons. (Light dark matter, for instance would produce
the former, whereas conventional heavy dark matter would generate the
latter via pion production.)

\subsection{Positron power from jets}

The jets in low/hard state or quiescent X-ray binaries are thought to
be mildly relativistic (see Gallo, Fender, Pooley 2003).  While there
are some arguments that the data on XRBs do allow for higher jet
velocities in the low/hard states (Heinz \& Merloni 2004), additional
indirect evidence supports the former assertion.  The giant jet
ejections that take place at transitions from hard to soft states are
well explained by a jet speed that increases as the state transition
proceeds, leading to a shock of the faster jet material against the
recently ejected slower material (Vadawale \etal 2003; Fender, Belloni
\& Gallo 2004).  Furthermore, the jets in low luminosity active
galactic nuclei are typically two-sided, while those in high
luminosity AGN are typically one-sided, consistent with the idea that
Doppler boosting is a more important effect in high luminosity AGN
than low luminosity ones.

We thus take a jet speed of 0.7c, meaning that the kinetic luminosity
of the jet is about $0.4\dot m c^2$.  This gives
$\dot{m}=3\times10^{14}$ g/sec for a jet power of about $10^{35}$
\ergs, typical in the quiescent systems whose X-ray luminosities are
about $10^{32}$ \ergs.  Theoretical studies suggest a range of
1000-10000 for the number of quiescent BH XRBs in the entire Galaxy
(Romani 1992; Portegies Zwart, Verbunt \& Ergma 1997).  From this
range, we derive an estimate of $10^{38-39}$ ergs/sec of total kinetic
power input into the Galactic interstellar medium (ISM) from quiescent
black hole XRBs.  Since roughly 1/3 of the stellar mass of the Milky
Way is in the Bulge, and low mass XRBs are good tracers of stellar
populations (Gilfanov 2004), this results in an estimate of $\sim$
300--3000 BH in the Bulge itself.  We can then estimate that the
kinetic power injected by quiescent BH X-ray binaries into the Bulge
is about $3\times10^{37}-3\times10^{38}$ ergs/s, and that the total
mass injected into the Bulge ISM from these jets is about
$10^{17}-10^{18}$ g/s, yielding a total proton injection rate of about
$5\times10^{40-41}$ per second.  If we see $\sim 10^{-3}$
photons/s/cm$^2$, then this means we see $10^{43}$ 511 keV photons per
sec.  If we assume that 3/4 of the annihilations go through the
3-photon positronium channel, then we need 2$\times$$10^{43}$
positrons/s produced to get the $10^{43}$/s 511 keV photons needed
(75\% produce no 511, 25\% produce two 511 photons).  We thus need a
pair to proton ratio of about 40-400 to account for the observed 511
keV luminosity.

The best attempts at estimating the pair fractions for jets come from
active galactic nuclei.  It has been shown by Sikora \& Madejski
(2000) that excessive pair fractions in quasar jets would lead to a
strong bump in the soft X-rays due to the bulk Compton upscattering of
the thermal photons from the accretion disk of the quasar, in
constrast to observations.  On the other hand, for lower luminosity
AGN, which are more likely to be analogous to the quiescent BH
X-ray binaries discussed here, the situation is less clear -- these
still could be pair-dominated as they do not have strong thermal
accretion disks to contribute photons for upscattering.  Comparisons
in M87 between the bulk kinetic power observed from jet-intracluster
medium interactions and the synchrotron luminosity argue that there
must be a large number of leptons per unit kinetic power, and hence
that the pair fraction must be large (Reynolds \etal 1996; Dunn,
Fabian \& Celotti 2006).  Reynolds \etal (2006) also find that the
pairs in these jets are most likely to be predominantly cold (i.e. the
energy spectrum for the electrons has no low energy cutoff, so that
the number density is dominated by electrons with low enough energies
to satisfy the constraint discussed in section 2.1, that high energy
pairs would produce excessive emission in $\sim$ MeV range via
in-flight annihilation).

Guessoum, Jean, \& Prantzos (2006) have estimated the production rate
of positrons in ``canonical'' jet-producing microquasars, based on the
energetics and models of these luminous XRBs which have been presented
in the literature; we summarize their results here.  They use an
average value of 10$^{41}$ pairs/sec from a luminous microquasar
producing ``steady'' jets (e.g. in the low/hard state, rather than the
large-scale episodic jets) at $L_X \sim$ 0.01-0.1$L_{Edd}$.  Using an
estimate of $\sim$100 for the total number of microquasars in the
Galaxy, together with the Galactic positional distribution of
observationally-confirmed microquasars, they derive an estimate of
$\sim$ 4.1$\times$10$^{42}$ positrons/s for the rate of annihilating
positrons in the Bulge.  They note that the positron production rate
they derive is smaller than what is inferred from the INTEGRAL 511 keV
observations.  Comparing their results with ours, we see that their
calculated positron injection rate from luminous microquasars within
the Bulge -- a fairly small population, of order $\sim$40 sources --
is an order of magnitude smaller than the $\sim 2\times10^{43}$
positrons/s we calculate as being produced by quiescent BH XRBs in the
Bulge -- a population we estimate to have $\sim$3000 members.  As
such, in the scenario we describe in this paper, the quiescent BH XRBs 
would be the dominant source of the Bulge 511 keV emission, although the 
luminous microquasars would also make a significant contribution.

From one year of INTEGRAL/SPI observations, Kn{\"o}dlseder \etal
(2005) found that the spatial distribution of 511 keV luminosity shows
a Bulge-to-disk (B/D) ratio of 3-9, which is higher than the mass of
the Galaxy in general.  Subsequently, using two years of INTEGRAL/SPI
data, Weidenspointner \etal (2007) revised this ratio downward to a
range of 1-4.  Since there exist mechanisms for producing 511 keV
emission in the Galactic Plane which involve young stars not present
in the Bulge, any viable mechanism for producing the bulk of the 511
keV emission in the Bulge must be one in which a larger amount of
annihilation per unit mass comes from the Bulge than from the Galactic
Plane.  At first glance, this seems to represent a problem for our
model, in which we suggest that 1/3 of the quiescent BH XRBs are
located in the Bulge; if the remaining 2/3 are simply assumed to be in
the Galactic disk, then the 511 keV luminosity B/D ratio would be 0.5,
well below the lower end of the range determined by Weidenspointner
\etal (2007).

However, this simple calculation neglects two important factors,
highlighted by Guessoum \etal (2006).  First, a significant portion of
the Galactic LMXB population is located in the Galactic halo, distinct
from the Bulge and disk LMXB populations.  Grimm \etal (2002) find
that 25\% of the total number of LMXBs in the Galaxy are located in
the halo.  They also find that 1/3 of bright LMXBs are located in the
Galactic Bulge, consistent with the expectation from the stellar mass
distribution of the Galaxy discussed above.  Thus if we assume that
the quiescent black hole XRBs described here have a similar
distribution to canonical luminous low mass XRBs, we expect $\sim$1/3
in the Bulge, $\sim$1/4 in the halo, and the remainder to be located
in the disk.  The halo sources are physically located well outside of
the Bulge and the Plane; thus they will not significantly contribute
to the 511 keV luminosity in the spatially-constrained Bulge and inner
disk areas (within $R \sim$ 1.5 kpc of the Galactic Centre) over which
the 511 keV emission has been detected (Weidenspointner \etal 2008).
Removing the halo sources results in a B/D luminosity ratio of 0.8 --
still below the lower limit of Weidenspointner \etal (2007).  (Note
that Guessoum \etal also made an estimate for the number of luminous
microquasars in the halo, and removed those sources before calculating
their Bulge-to-disk luminosity ratio.)

The second factor, as discussed by Kn{\"o}dlseder \etal (2005), is
that the scale height of XRBs in the Galactic disk is about 700 pc
(Jonker \& Nelemans 2004), considerably larger than that of the gas in
the Galaxy, and that this can have profound effects on the inferred
rate of positron production in different parts of the Galaxy.  If
positrons are injected into gas poor parts of the Galaxy, they may
travel fairly large distances before annihilating, yielding a
situation where the distribution of locations for positron
annihilation is not the same as the distribution of locations for
positron production. The largest effect would be to transfer positrons
from production sites several gas scale heights above the Galactic
Plane into the halo or Bulge before they undergo annihilation.
Following Guessoum \etal (2006), if we estimate that 50\% of positrons
produced in the disk escape into the halo and/or propagate along field
lines towards the Bulge, while all positrons produced in the Bulge 
are retained and annihilate therein (see \eg Jean \etal 2006), then we
must reduce the disk 511 keV luminosity by 1/2.  Combining this factor
with the Bulge and disk population estimates above, we now find a B/D
511 keV luminosity ratio for the quiescent BH XRB population of
$\sim$1.6, well within the range for the B/D ratio found by
Weidenspointner \etal (2007).  Finally, we note that some positrons
which escape from the disk may end up annihilating in the Bulge,
rather than in the halo, resulting in a still larger B/D ratio.

Therefore, it seems plausible that the 511 keV emission could come
from the jets of quiescent (X-ray faint) black hole X-ray binaries.
The requirements to make this happen -- {\it (i)} that the jets have
$\sim100$ positrons per proton; {\it (ii)} that there are on the order
of 3000 quiescent BH X-ray binaries in the Bulge; and {\it (iii)} that
the pairs are mostly cold -- are all within the range of reasonably
standard assumptions and well within the range of observational
constraints.

\subsection{Predicted Emission at Other Wavelengths}

We first introduce a few standard assumptions about jet kinetic power
from accreting binaries, and its relationships with observables such
as radio and X-ray luminosities.  The kinetic power input into a
relativistic jet is assumed to be a constant fraction of the accretion
power:
\begin{equation}
L_K = f\dot{M}.
\end{equation}
This follows in a straightforward manner from the standard
mechanisms for producing jets by extracting either the spin energy of
the central black hole (Blandford \& Znajek 1977), or the rotational
energy of the accretion disk (Blandford \& Payne 1982), provided that
the height to radius relation for the accretion disk does not change
as a function of accretion rate (\eg Livio, Ogilvie \& Pringle 1999;
Meier 2001).  

Based on the results of standard synchrotron theory
(\eg Blandford \& K\"onigl 1979; Falcke \& Biermann 1995; Heinz \&
Sunyaev 2003), it is normally assumed that the observed radio
luminosity from a jet is proportional to its kinetic luminosity to the
1.4 power:
\begin{equation}
L_R \propto L_K^{1.4}
\end{equation}

Next, it is assumed that black holes at low accretion rates (in the
hard or quiescent states) are radiatively inefficient, because they
can advect energy across their event horizons, leading to a scaling of
the bolometric luminosity on the square of the mass accretion rate
(Narayan \& Yi 1994)
\begin{equation}
L_{X,bh} \propto \dot{M}^2;
\end{equation}
whereas neutron stars, with
their solid surfaces, cannot advect matter or energy, leading to a
linear proportionality between mass transfer rate and luminosity:
\begin{equation}
L_{X,ns} \propto \dot{M}.
\end{equation}

The resulting expectations are that black holes should have a radio
luminosity which depends on the X-ray luminosity to the 0.7 power,
assuming that the radio emission comes from the jet, and the X-ray
emission comes from an advection-dominated accretion flow:
\begin{equation}
L_{R} \propto L_K^{1.4} \propto \dot{M}^{1.4} \propto L_{X,bh}^{0.7}.
\end{equation}
Similarly, neutron stars should have a radio luminosity which scales
as the X-ray luminosity to the 1.4 power.  Both these correlations are
indeed observed; see \eg Corbel \etal (2003) and Gallo, Fender \&
Pooley (2003) for the case of black holes, and Migliari and Fender
(2006) for the case of neutron stars.

The kinetic energy input into the ISM from black hole X-ray transients
in quiescence can then be estimated based on their X-ray luminosities.
The characteristic X-ray luminosities of quiescent BH X-ray transients
are $10^{32}$ ergs/sec.  Jets are likely to take away a substantial
fraction of kinetic power in the low/hard state (\eg Malzac, Merloni
\& Fabian 2004), but, given that state transitions between the
radiatively efficient high/soft state in which little or no jet power
is seen, and the radiatively inefficient low/hard state in which there
is a steady jet, do not correspond with abrupt luminosity changes, the
jet power cannot be much greater than the radiated power at the state
transition (Maccarone 2005b).  Therefore, the best estimate of the jet
kinetic power is that it scales with $L_X^{0.5}$, as predicted by the
model, with the normalization fixed such that the jet power and the
X-ray luminosity are equal at the state transition luminosity which is
typically about 2\% of the Eddington luminosity (Maccarone 2003).
Such arguments have been used to estimate that the total kinetic power
input into the ISM from {\it bright} XRBs is about $10^{39}$ ergs/sec
(Fender, Maccarone \& van Kesteren 2005; Heinz \& Grimm 2006), while
the kinetic power input from {\it quiescent} black hole X-ray binaries
may be $10^{38-39}$ ergs/sec (as derived in Section 3.2; see also
Maccarone 2005a).  (Note that we use this power estimate based on the
theoretical numbers of the BH XRB population rather than using the
luminosity function derived from observed bright BH X-ray transients,
as the latter option would involve extrapolating the luminosity
function to X-ray luminosities 3--6 orders of magnitude lower than
what has been directly measured for the bright sources.)  Since the
total kinetic power in jets from quiescent BH X-ray binaries is likely
to be of the same order as that from the most luminous systems, and
the quiescent systems will be more strongly concentrated in the Bulge
than will the luminous XRBs (a population which includes HMXBs as well
as LMXBs), it is to be expected that a substantial fraction of the
Bulge positrons come from quiescent black hole X-ray binaries.

\subsection{Observational Constraints at Other Wavelengths}

Finally, we note two other diffuse components in the Galactic Centre
that may be related to the annihilation radiation.\\

\noindent {\it X-ray: } \\

In the bulge core, diffuse Chandra emission amounts to $2 \times
10^{36}$ergs/s, with a spectrum equivalent to an 8 keV thermal plasma
over the 17x17 sq. arcmin Chandra field (Muno \etal 2004).  The
observed spectral shape is indicative of a truly diffuse component
rather than originating from a collection of unresolved discrete
sources.  The flux limit of the Muno \etal (2003) X-ray survey is
$\sim$10$^{31}$ ergs/s.  Normal stars have luminosities of $\sim
10^{27}$ergs/s or less, and so at least $10^9$ normal stars would be
needed to account for the diffuse emission component.  However, $10^9$
stars are all the (solar mass) stars in the Bulge, not just those
within the central 40 square parsecs covered by the Muno \etal Chandra
survey.  It has been argued that this plasma can be thermally
maintained by low energy cosmic ray heating (Yusef-Zadeh \etal 2007).
Could these cosmic rays possibly come from the X-ray binary jets
themselves?  The energetics should be adequate.  One question is
whether this fraction of the energy actually comes out in cosmic rays,
rather than just leading to bulk
kinetic motions of the gas at the collisional interface of the jets.\\

\noindent {\it Microwave:}\\

WMAP finds a likely synchrotron background with an integrated
luminosity of about 2$\times10^{36}$ ergs/sec in the 23-61 GHz
frequency range from a region 20--30$^\circ$ in radius around the
Galactic Centre region (Finkbeiner 2004a).  The total power emitted at
these frequencies is estimated to be between 1--5$\times 10^{36}$ erg
s$^{-1}$.  Assuming a distance of 8.5 kpc, this corresponds to a flux
level between 3000 and 15000 Jy.  We note that there are large
uncertainties in this estimate, given the need to subtract many other
contributions, including thermal dust, spinning dust, thermal
bremsstrahlung emission and synchrotron from electrons accelerated in
SNe shocks, from the microwave maps.  Moreover, use of the WMAP
internal linear combination (ILC) dust extinction template leaves much
uncertainty in possible spatial variations of the
composition-dependent frequency sensitivity of the dust emissivity.
Nonetheless, the signal has generated interest given its possible
connection to exotic processes such as the annihilation of massive
WIMPs (Finkbeiner 2004b; Hooper, Finkbeiner \& Dobler 2007), or states
of `excited' dark matter (Finkbeiner \& Weiner 2007).

The reported haze flux level cannot be produced from core emission
from individual black hole X-ray binaries -- as previously discussed,
no more than about 3000 quiescent BH XRBs are thought to be in the
entire GC region (\eg Romani 1992; Portegies Zwart, Verbunt \& Ergma
1997), meaning that the flux of the typical source would have to be
$\sim$1 to several Jy.  Sources above even 100 mJy would be well-known
radio sources at lower frequencies (since they would likely be flat
spectrum sources, or sources whose flux density drops with increasing
frequency), and an excess of such sources would be clear from existing
radio surveys.  The typical flux of a quiescent low mass X-ray binary
in the radio at the Galactic Centre distance should be $\approx$
10--20 $\mu$Jy (see \eg Gallo \etal 2006 for the faintest known system
of this sort; Hynes \etal 2004 for V404 Cyg, the brightest one).
Given $\sim$3000 BH with an average quiescent radio flux level of
$\sim$100 $\mu$Jy, the integrated luminosity of those outflows at WMAP
wavelengths would be far less than that of the WMAP GC haze (even if
the haze fluz has been significantly overestimated).  This may support
the interpretation of the haze emission source as relativistic
electrons, produced by annihilating 100 GeV dark matter particles
(Finkbeiner 2004b; Hooper, Finkbeiner \& Dobler 2007).

\section{Discussion}

It thus seems reasonable that most or all of the Galactic Centre 511
keV emission line could come from low luminosity BH X-ray binaries.
We note that this scenario is not required, since the
positron-to-proton ratios for X-ray binary jets are largely
unconstrained by observations.  What about the contribution from high
luminosity (``canonical'') XRBs?  We need to estimate the fraction of
the overall 511 keV line flux that can be accounted for by the
well-studied canonical jet sources, and correspondingly the percentage
of the flux which needs to be accounted for by the low luminosity
population.  However, if we only consider black hole XRBs, since (as
previously discussed) the NS jet flux is too small, then at any given
time there are only a few active luminous hard state black holes which
would contribute to the 511 keV emission.  But in this case, if a
significant fraction of the overall 511 keV flux originated from a
handful of sources, and if the positrons do not propagate and
annihilate far from their production sites, we would expect to see
these individual sources in the 511 keV flux maps - the 511 keV line
emission would not appear to be as smooth as it does.

\begin{figure*}
\centering
\subfigure[LMXB luminosity function]
{
\label{fig:sub:lumf1}
\includegraphics[width=0.45\textwidth,  keepaspectratio]{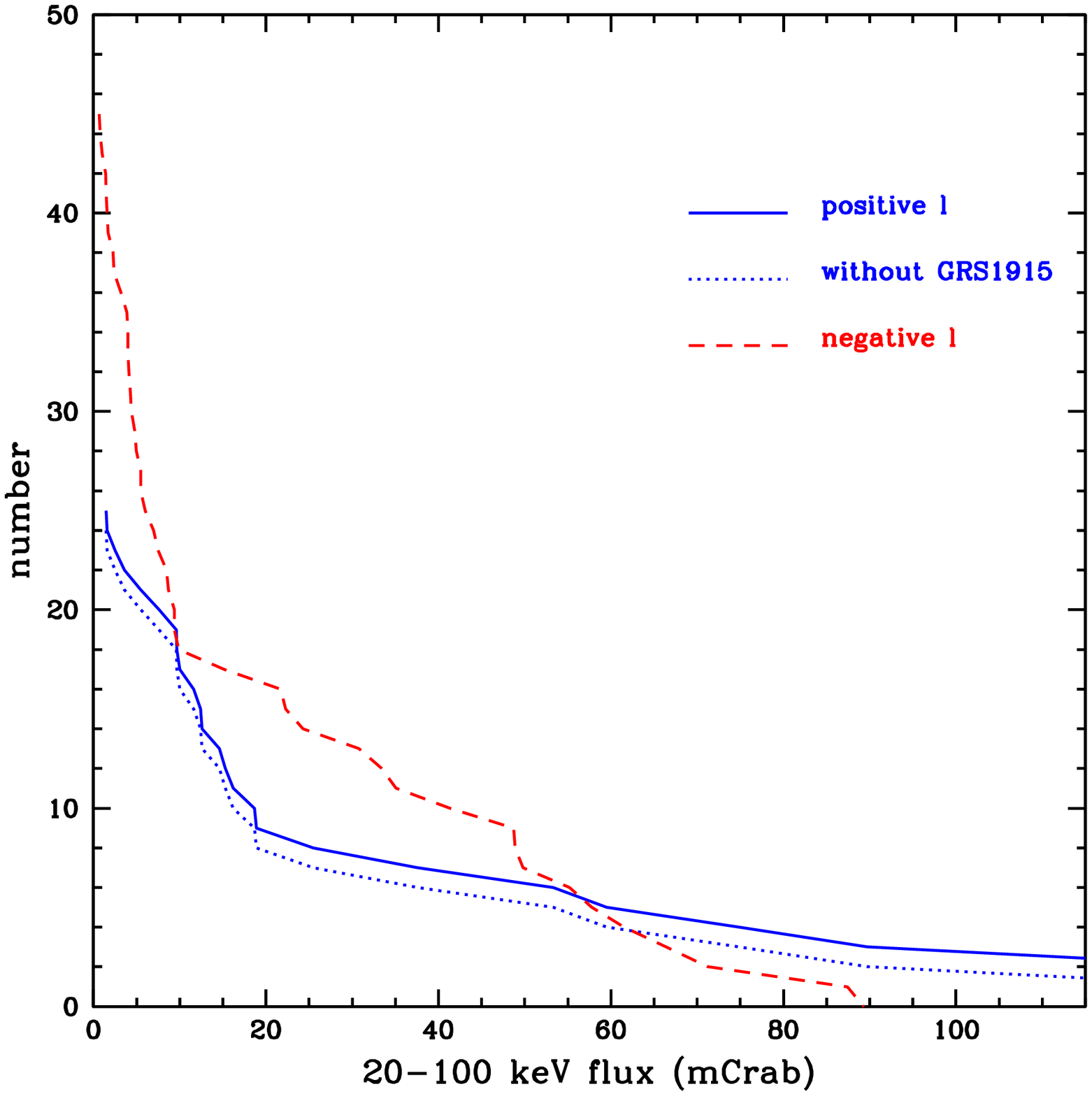}
}
\subfigure[Integrated flux] 
{
\label{fig:sub:lumf2}
\includegraphics[width=0.45\textwidth,  keepaspectratio]{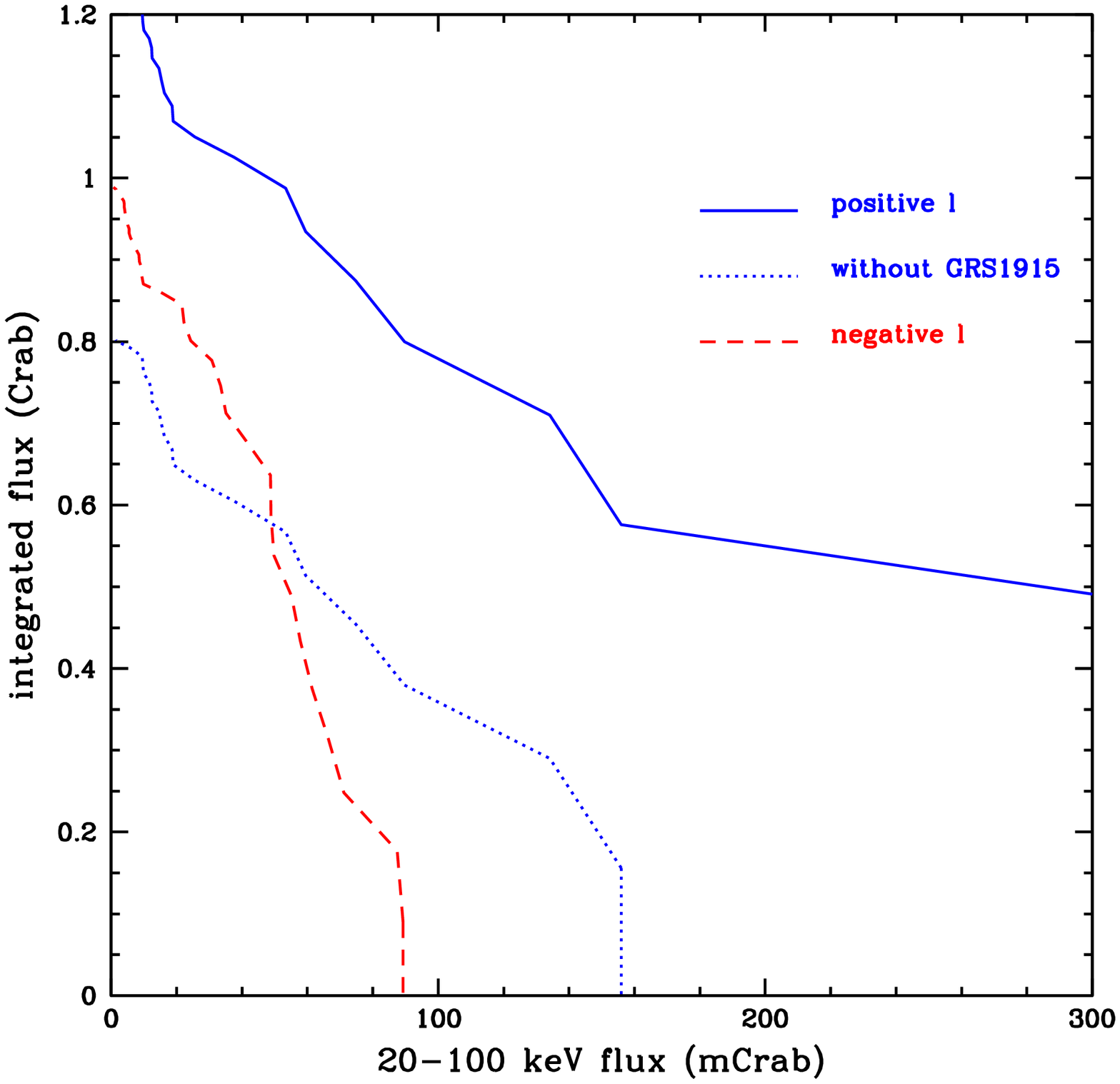}
}
\caption{(a) Cumulative 20--100 keV luminosity function for all
Galactic LMXBs within $\pm$10\degs~ latitude of the Plane, as reported
in the catalogue of Bird \etal (2007).  The solid line is for {\it all}
LMXBs at positive longitudes, the dotted line is for LMXBs at
positive longitudes {\it excluding} GRS~1915+105 (by far the most
luminous source in the catalogue), and the dashed line is for LMXBs
at negative longitudes.  (b) Integrated flux contributed by LMXBs
over a given flux limit for the same samples of sources as in panel
(a); line styles are as in panel (a).}
\end{figure*}

There is no convincing evidence for time variability of the 511 keV
line.  It is clear that the mechanism proposed here -- summed emission
from hundreds to thousands of sources of roughly equal importance --
should predict low variability levels in the 511 keV emission.  It is
less clear whether this would be true from many of the other
mechanisms proposed which rely more heavily on a smaller number of
sources.  However, temporal variability is not useful for
distinguishing the total number of sources if the characteristic
positron lifetime is $10^5$ years as suggested by Ferrer \& Vachaspati
(2005).  Only spatial variability would be useful in this case.

Weidenspointner \etal (2008) suggest that an asymmetry in the 511 keV
emission is correlated with an asymmetry in the LMXB distribution
around the inner Galaxy.  The asymmetry argument is that the disk 511
keV flux is asymmetric, and so is the hard LMXB population, with the
same ratio between halves (positive and negative galactic longitudes)
of the disk.  Furthermore, the number of objects ($\sim$70) times a
plausible mean flux per source ($10^{-5}$ photons/cm$^2$/s each) gives
the right total disk flux ($7\times10^{-4}$ photons/cm$^2$/s); thus
the luminous hard LMXBs may be the source of the disk flux, if not the
($\sim10^{-3}$ photons/cm$^2$/s) Bulge flux (note that the latter
corresponds to a larger luminosity, since the Bulge is further away;
overall it should be 3-9 times the disk luminosity).

While the positional asymmetries in the known luminous X-ray binary
distribution do appear to be in the same sense as the asymmetry in the
511 keV emission, it is not clear that the asymmetry in the Bulge
component of the X-ray binary population is real.  The inner 20\degs
of the Galaxy show only a 1.5$\sigma$ asymmetry - 30 (negative
longitude) versus 18 (positive longitude) sources.  In other words,
given a total of 48 sources, the probability of having $\geq$30
sources at negative latitudes is 5.6\%.  We note that at least eight
of the X-ray binaries in the INTEGRAL catalog (5 at negative
longitudes, 3 at positive) used by Weidenspointner \etal (2008) in
their determination of the asymmetry are ``foreground'' sources, with
known distances which are not consistent with being in the Galactic
Bulge (Jonker \& Nelemans 2004).  If the 511 keV emission is
physically associated with the population of hard XRBs, as is
suggested by the asymmetry, then the observed enhancement of 511 keV
flux in the central region of the Galactic Plane is simply a
projection effect along our line of sight towards the GC, which
coincidentally results in an apparent concentration of 511 keV line
emission within $l$\pms$\sim20$\degs.  Thus in this scenario, the 511
keV enhancement is not physically associated with the Galactic Bulge
itself, and the morphology and intensity of the 511 keV map is
dependent upon the longitudinal location of bright, hard XRBs.
However, we note that GRS~1915+105, which is almost certainly the
strongest emitter of positrons from jets outside of the innermost
region of the Bulge, is on the other side of the Galaxy (at positive
galactic longitude) from the bulk of the INTEGRAL sources and the
observed 511 keV enhancement (which are at negative galactic
longitudes).

The asymmetry interpretation has at least 4 distinct problems:
\begin{enumerate}
\item The LMXB luminosity function is steep near the lower end;
changing the sensitivity limit of the IBIS catalog would change the
positive/negative longitude number ratio.
\item Integrating flux down the luminosity function, the asymmetry
also depends on the catalogue limit, in that the total flux asymmetry
is actually the reverse of the number asymmetry, at least at high
luminosities.
\item A few sources (e.g. GRS~1915+105) are responsible for much of
the flux; if this is always true, then the numbers of objects in the
catalog may not mean much; a few bright sources could determine how
many positrons are produced at any given time (i.e. the total flux
need not correlate with the number of sources over the IBIS limit).
\item Since LMXB X-ray emission is time-variable, there is no
indication that any current asymmetry in numbers or flux would be
constant in time.
\end{enumerate}

Fig.\ 1a shows the cumulative (20-100 keV) luminosity function for all
72 Galactic LMXBs within $\pm$10\degs\ latitude of the Plane in the
catalogue of Bird \etal (2007), divided up into objects at positive
and at negative longitudes (solid and dashed lines respectively).
Note this scale height excludes high-latitude luminous LMXBs such as
Sco X-1.  Fig.\ 1b shows the integrated flux contributed by binaries
over a given flux limit for each of these samples.

Clearly the longitudinal asymmetry in number counts is sensitive to
the lower flux limit of the catalogue; here, this is set to the
INTEGRAL/IBIS 20-100 keV detection limit.  The current integrated flux
is dominated by a small number of bright sources, with the majority of
the sources in the INTEGRAL catalogue being just at the flux detection
limit.  Given the recently identified large population of X-ray faint
XRBs discussed above, it is clear that there are many lower luminosity
sources below the INTEGRAL detection threshold.  The asymmetry in the
source number counts could thus easily be an artifact of the flux bias
of the INTEGRAL sample.  Therefore the source numbers used to
determine the positional asymmetry could be greatly changed by a
slight change in the flux cutoff (to either higher or lower
luminosities).  Cutting objects below 10 mCrab, for instance, would
produce no asymmetry at all, while cutting at a high flux limit would
produce more sources at positive longitudes.  So overall, the observed
positional asymmetry seen in the INTEGRAL catalogue, with its
instrumentally-determined bias towards luminous sources, may not be
significant even if the observed current 20-100 keV flux were to
correlate exactly with the 511 keV flux.

Furthermore, the integrated flux from LMXBs in the catalogue is
dominated by objects at positive longitudes at all flux limits (Fig.\
1b), although admittedly this is mainly due to the strong contribution
from the single brightest source, GRS~1915+105.  Thus the asymmetry in
the number of sources does not agree with the asymmetry in the total
flux from the INTEGRAL catalogue sources -- there is more integrated
flux where there are fewer sources (positive longitudes), and less
integrated flux where there are more sources (negative longitudes).

The fact that the current ``snapshot'' distribution of the integrated
flux does not match the 511 keV asymmetry may in itself not be
problematic for the Weidenspointner \etal (2008) hypothesis, since the
timescale for positrons to annihilate and produce 511 keV emission is
$\simgt 10^5$ years.  However, we note that a large fraction of the
sources observed by INTEGRAL and used to ``find'' the positional
distribution asymmetry are transients.  In addition, those X-ray
binaries which we call ``persistent'' merely means that the sources
have been consistently luminous over the $\sim$35 years of
observational X-ray astronomy, which provides very little information
about the persistence of their luminosity over the $10^5$ year
timescale of photon interaction.  Thus the current apparent positional
asymmetry is merely a snapshot in time, and does not constitute
evidence that the total absolute numbers of 511 keV-producing binaries
in the direction of the inner Galaxy -- including those currently in
quiescence, which are likely to be the majority -- at negative
longitudes is larger than at positive longitudes.

Over the long lifetime of positron interaction, the integrated flux from
all sources which become active during that time may indeed be larger
at negative longitudes than positive ones, resulting in the observed
asymmetry in the 511 keV emission.  To determine this, however, we
would need to be certain that there is a true asymmetry in the
distribution of all positron-producing XRBs in the inner Galaxy - not
just during the current epoch, when only a small fraction of such XRBs
are observed to be active and luminous.  We would have to assume that
the asymmetric distribution of currently luminous XRBs - a sample of
$\sim$70 sources - reflects an underlying asymmetry in the overall
distribution of all XRBs - on the order of $\sim$3000 BH binaries -
which would contribute to the 511 keV emission over $10^{5}$ years.
While this is certainly possible, it cannot be considered conclusive,
as it is dependent upon both the arbitrary lower flux cutoff of the
INTEGRAL catalogue (set by the instrument sensitivity rather than by
any intrinsic property of the sources themselves) and the arbitrary
time of observation (\eg the modern era of high-energy astronomy).

A much deeper census of the black hole X-ray binary population, which
explores down to the quiescent luminosity of these sources ($\simlt
10^{32}$ erg/s), is needed in order to search for a true asymmetry in
the physical distribution of jet-emitting XRBs in the inner Galaxy.
Such a census would require both a deep X-ray survey and infrared
follow-up observations to identify stellar counterparts to a
representative sample of the X-ray sources, in order to conclusively
determine which sources contain black holes.  It would then be
possible to assess whether or not the asymmetry in the 511 keV flux is
due to an intrinsic spatial asymmetry in the number of XRBs which
episodically exhibit large X-ray luminosities integrated over the
timescale for 511 keV production.

If stellar sources do account for 511 keV emission, it is unlikely
that they would show up as point sources at higher spatial resolution
at other wavelengths than the X-ray, as the positrons likely propagate
a significant distance ($\sim$50-250 pc, which is $\sim$0.5\degs --
1.5\degs at the GC; Jean \etal 2006 -- but see also Prantzos \etal
2006, which suggested a larger distance) before annihilating.  For the
brightest of the low-luminosity sources ($10^{35}$ erg/s), we could
perhaps detect a flat near-IR spectrum which would not be consistent
with the flux of a mass donor star alone, but which instead would
denote the presence of another component - \eg a jet outflow. It is
not clear if there is enough sensitivity in the mid-IR for
ground-based observations \eg at 4 and 8 microns to detect the flux
from these sources.  Similarly, for sources with this X-ray
luminosity, high frequency (8.6 GHz) radio emission might be
detectable (\eg with the VLA) to search for the flat spectrum
signature associated with the steady jets from hard state X-ray
binaries.

\section*{Acknowledgements}

The authors wish to thank Tony Bird, Sera Markoff, and Philipp
Podsiadlowski for helpful discussions.  We also thank Mark Rayner for
his preliminary work on extracting OSSE observations of X-ray binaries
from the HEASARC archives.  We thank the referee for a close
reading of the manuscript and useful suggestions.  JET gratefully
acknowledges support from PPARC, the Aspen Center for Physics, the US
DoE (grant AST-0307859), the US NSF (contract DE-FG02-04ER41316), and
NSERC Canada.

\onecolumn

\end{document}